# Measurements of positive ions and air-earth current density at Maitri, Antarctica

Devendraa Siingh*, Vimlesh Pant and A K Kamra

Indian Institute of Tropical Meteorology, Pune, India

**Abstract**

Simultaneous measurements of the small-, intermediate- and large- positive ions and air-earth current density made at a coastal station, Maitri (70° 45' 52" S, 11° 44' 03" E, 130 m above sea level) at Antarctica during January – February 2005, are reported. Although, small and large positive ion concentrations do not show any systematic diurnal variations, variations in them are almost similar to each other. On the other hand, variations in intermediate positive ion concentrations are independent of variations in the small/large positive ions and exhibit a diurnal variation which is similar to that in atmospheric temperature on fair weather days with a maximum during the day and minimum during the night hours. No such diurnal variation in intermediate positive ion concentration is observed on cloudy days when variations in them are also similar to those in small/large positive ion concentrations. Magnitude of diurnal variation in intermediate positive ion concentration on fair weather days increases with the lowering of atmospheric temperature in this season. Scavenging of ions by snowfall and trapping of α-rays from the ground radioactivity by a thin layer of snow on ground, is demonstrated from observations. Variations in intermediate positive ion concentration are explained on the basis of the formation of new particles by the photolytic nucleation process.

***Present address:** Institute of Environmental Physics, University of Tartu, 18, Ulikooli Street, Tartu- 50090, Estonia.

## 1. Introduction

Electrical state of the atmosphere is mostly determined by atmospheric ions. Their movements under electrical/non-electrical forces cause the flow of conduction and convection currents that determine their spatial and temporal distributions at any place. Concentrations and mobilities of these ions range over four orders of magnitude and are generally governed by the processes which generate and destroy them. The aerosol particles and trace gases ubiquously present in the atmosphere play a major role in their evolution. Temporal and spatial variations of ions have been studied at several places [Wait and Torreson, 1934; Norinder and Siksna, 1953; Misaki, 1961; Jonassen and Wilkening, 1965; Misaki et al., 1972]. Dhanorkar and Kamra, [1991, 92, 93] have extensively studied the diurnal and seasonal variations of different categories of ions and their relative contributions to the atmospheric electric conductivity at a tropical land station, Pune. Horrak et al. [2000, 2003] categorize the atmospheric ions in five different categories and study their properties at a mid-latitude station, Tahkuse, Estonia. Hirsikko et al. [2005], from their measurements made at a station in southern Finland, observe that the concentrations and growth rates of the atmospheric ions and particles displayed a seasonal variability.

Concentration of atmospheric ions over land areas is determined by several sources and sinks. Further, characteristics of these ions on land areas in tropical and mid-latitude regions are grossly modified by the naturally and anthropogenically produced pollutants. So, their contributions to atmospheric electric conductivity are complex. However, at clean places such as over open oceans and polar regions, contributions of different types of ions to the atmospheric electric conductivity is mainly determined by the rate of small ion generation and concentration of background aerosols. The inverse relationship observed between the atmospheric electric conductivity and aerosol concentration has therefore been often used as an index of background air pollution at such places [Cobb and Wells, 1970; Kamra and Deshpande, 1995, Deshpande and Kamra, 2002].

Antarctica provides a unique site which is practically free of anthropogenic pollution. Since more than 98% of the continent is covered by ice, the ionization produced by the ground radioactivity and its emissions which is a major source of ionization near the ground, is almost nil. However, being located near the South Magnetic Pole and as a result, the lines of force of Earth's magnetic field being perpendicular to the Earth's surface. The continent is exposed to a greater flux of high energy particles which are known to penetrate and produce more ionization in the lower atmosphere than at other sites at lower latitudes. Therefore, the sources and sinks of ions and their transport in the Antarctic environment are much different than those in the tropics and mid-latitudes. Although measurements of several atmospheric electric parameters have often been made (e.g. Kasmier, 1972; Byrne et al., 1993; Burns et al., 1995; Bering et al., 1998; Deshpande and Kamra, 2001; Virkkula et al., 2005), to the author's knowledge, no measurements of atmospheric ions of different categories have so far been made at Antarctica. Knowledge of atmospheric ions is important not only to understand the electrical state of the atmosphere but also in understanding the global electric circuit [Roble and Tzur, 1986; Tinsely, 2000; Rycroft et al., 2000; Singh et al., 2004; Siingh et al., 2007], nucleation and growth characteristics of aerosol particles [Tinsley and Heelis, 1993; Carslaw et al., 2002; Harrison and Carslaw, 2003; Nadykto and Yu, 2003; Hirsikko et al., 2005], monitoring of background air pollution [Cobb and Wells, 1970; Kamra and Deshpande, 1995], and in understanding the solar-terrestrial relationships [Markson, 1978; Markson and Muir, 1980].

Here, we present our simultaneous measurements of the small-, intermediate-, and large- positive ions and air-earth current density made at Maitri, (70° 45' 52" S, 11° 44' 03" E, 130 m above mean sea level) Antarctica in January - February, 2005 during the 24$^{th}$ Indian Scientific Expedition to Antarctica.

## 2. Instrumentation and Measurement Site

Positive ion concentrations were measured with an ion-counter of the type described by Dhanorkar and Kamra [1991] and shown in Figure 1. It consists of three coaxial Gerdiens condensers through which the air is sucked with a single fan. Table 1 details the measurements of all the three condensers. Ranges of mobility covered by each condenser are described in Table 2.

Several steps were taken to minimize the effects of the harsh Antarctic weather on our measurements. Non-magnetic stainless steel was used for fabricating the ion counter. All electronic components used for making the electronic circuitry were of military grade which can operate well up to subfreezing temperatures of - 40° C. Coaxial cables with teflon insulation were used for carrying signals from ion counter to the data-logger. To minimize the effect of strong winds on the suction rates in condensers, ion counter was placed with its condensers perpendicular to the prevailing wind direction.

To check the stability of the instrument, zero-checking of individual condensers were performed periodically by grounding the inputs. No appreciable zero-shift was observed in fair weather and even in light snowfall. Whenever the snowfall continued beyond a few hours, zero was found to shift and observations were discontinued.

Air-earth current density is measured with a 1 $m^2$ flat-plate antenna kept flush with the ground and is shown in Figure 2. The inputs from all the three condensers of the ion counter and the air-earth current plate are amplified with separate amplifiers placed close to the sensors and then fed through coaxial cables to a data-logger placed in a nearby hut.

Measurements were made at Maitri located in the Schirmacher oasis in the Dronning Maud Land, East Antarctica. The east trending oasis is exposed over an area of 35 km$^2$ with 16 km length and a maximum of 27 km width. It has a lake and then steep cliffs towards the ice-shelf and is covered by polar ice on the southern side. The area is dominantly covered by sandy and loamy sand type of soil. Measurements of Rn$^{222}$ and small ions made at Maitri show that their atmospheric concentrations close to ground are very low and not much different from that over sea [Ramachandran and Balani, 1995].

Figure 3 shows the location of Maitri (70° 45' 52" S, 11° 44' 03" E, 130 m above sea level) at the Antarctic continent, the location of instruments at the Maitri station and a scatter polar diagram of 3-hourly surface winds for the period of observations at Maitri. Maitri is located in ~ 80 km south of the coastal line. Prevailing direction of winds is southeasterly. Ion counter was placed on the ground with the inlet of the small and intermediate positive ion condensers at 50 cm and large ion condenser at 60 cm above the ground. The air-earth current plate was placed flush with the ground. The single storeyed building of Maitri station, the generators, gas plant and incinerator are about ~ 300 m away in the southwest direction from the instruments so that any pollutants released from them has no or little chance of reaching the site of measurements with the prevailing winds.

## 3. Observations

### 3.1. Daily average concentrations of ions

Figure 4 shows the daily average values of the small-, intermediate-, and large positive ion concentrations and air-earth current density for the entire period of measurements at Maitri. Vertical bars show the standard deviations. Concentrations of small, intermediate and large positive ions vary in the ranges of 2 to 6 x $10^2$ $cm^{-3}$, 7 x $10^2$ to 3 x $10^3$ $cm^{-3}$, and from 5 x $10^3$ to 1.2 x $10^4$ $cm^{-3}$ respectively and the air-earth current density varies from 0.5 to 1.6 x $10^{-12}$ A $m^{-2}$ during this period. Values of all categories of ions start increasing between the Julian day 15 to 18 and, except for a dip observed around Julian day 22, are 1.5 to 2 times larger up to the Julian day 45 as compared to the periods immediately before or after this period. Variations in air-earth current during this period follow similar trend. In comparison to continental stations in the northern hemisphere, these values are roughly 50 – 80% lower than that at a tropical station, Pune [Dhanorkar and Kamra, 1992] and more than those at mid-latitude station at Tahkuse [Horrak et al., 2003] in the northern hemisphere. In particular, the intermediate positive ion concentration at Maitri is about an order of magnitude higher than those at Tahkuse in this season.

The variations in electrical parameters don't show much similarity with the variations in meteorological parameters at Maitri (Figure 5). The daily average atmospheric temperature, however, from Julian day 15 to 47 drops by 4°C, from 2°C above freezing point to 2°C below freezing point. Very strong winds prevailed after 1800 UT and the station experienced snowfall during the night of Julian day 46.

A characteristic feature in our measurements is 2 – 4 hours periods when the small and large positive ion concentrations simultaneously increase or decrease by about an order of magnitude. Air-earth current density also shows a similar change during these periods. However, intermediate positive ion concentration does not necessarily show any corresponding change. Figure 6 shows two examples of such changes when the intermediate positive ion concentration does not show (a) or shows (b) changes similar to, small/large positive ion concentrations. Such changes are not associated with change in any of the 3-hourly observed meteorological parameters at Maitri. Further, an examination of the 5-day back-trajectories around and during such events does not show any significant change in the place of origin of the air mass. However, association of such changes with changes in the high energy particle fluxes which may cause change in ionization and thus increase the small ion concentrations, can not be ruled out. The corresponding changes in large positive ion concentration may follow due to the ion-aerosol attachment process.

**3.2. Diurnal variations**

At Maitri, concentrations of small and large positive ions do not show any systematic diurnal variations on fair weather days. However, they exhibit a very high degree of similarity in variations amongst themselves. For example, Figure 7 shows diurnal variations of the concentrations of three categories of positive ions and the air-earth current density on a typical fair weather day, January 10, 2005. On this day, atmospheric pressure varied from 971 to 973 hPa with no trace of cloud throughout the day, except for moderately strong southeasterly winds during night and morning hours. Winds were calm or very light and temperature gradually increased from -1°C at 0100 UT, attained a maximum of 6°C at 0900 UT and then decreased to 2°C at 2400 UT. Sunshine lasted for almost all the 24 hours of the day. Small positive ion concentrations on this day were less than 3 x $10^2$ cm$^{-3}$ and were minimum at about 0900 UT. Large positive ion concentrations, though more than an order of magnitude higher, closely follow the variations in small positive ion concentration. The intermediate positive ion concentrations change little in the range of 1 to 2 x $10^3$ cm$^{-3}$ and vary independent of variations in small/large positive ion concentrations. In contrast to the small/large positive ion concentrations, the diurnal variation in intermediate positive ion concentrations are much regular and generally follow the trend in atmospheric temperature and are comparatively higher during the day than the night hours. The air-earth current density varies from 0 to 1.8 x $10^{-12}$ A m$^{-2}$ and almost exactly follows the variations in small/large positive ion concentrations.

During afternoons, the values of small and large positive ion concentrations at Maitri are comparable to those observed at tropical station, Pune [Dhanorkar and Kamra, 1993] and at mid-latitude station, Tahkuse [Horrak et al., 2003]. However, the maxima observed in the morning hours at Pune are not observed at Maitri. The day and night values also don't show much systematic difference at Maitri as at Pune and Tahkuse. On the other hand, the intermediate positive ion concentrations at Maitri are about an order of magnitude higher than that at Tahkuse. The intermediate positive ion concentrations at Maitri are comparable to the afternoon values but much lower than the morning maxima values at Pune. Diurnal variation of intermediate ion concentrations shows a distinct maximum at all the three locations but at different times e.g. during 0600-0800 UT at Pune, during 1200 to 1300 UT at Tahkuse and during 0900-1600 UT at Maitri. Unlike at Pune ot Tahkuse the maxima at Maitri is much broader and is almost flat for several hours in the afternoon.

In sharp contrast to the diurnal variations on fair-weather days, the intermediate ion concentrations on cloudy days do not show any maximum in the afternoon hours and follow the variations in small/large positive ion concentrations (Figure 8). For example, on January 18, 2005, the sky remained covered for the entire day with more than 6 octa of clouds. Southeasterly winds of 2 – 5 ms$^{-1}$ and temperatures from - 2 to + 3°C prevailed over the station. Concentrations of all categories of positive ions were comparatively small and showed much smaller variability during the whole day. Variations in air-earth current density on such cloudy days closely follow the variations in either category of positive ions.

### 3.3. Effect of snowfall

Snowfall, just like rainfall, is considered to be an effective scavenger of atmospheric aerosols and ions. Our observations at Maitri provide several good cases of its demonstration. For example, Figure 9 shows that with the start of snowfall at about 0400 UT on the Julian day 22 the concentrations of all categories of ions fall by about an order of magnitude and continue to be low until 0800 UT when observations had to be discontinued due to heavy snowfall. During this period, sky remained overcast with southeasterly winds remaining below 5 ms$^{-1}$ and atmospheric temperatures in the range of 0 to – 2°C. Lowest values of the ion concentrations encountered during this period are 10 ions cm$^{-3}$ for small positive ions, $10^2$ ions cm$^{-3}$ for intermediate positive ions and 3 x $10^2$ ions cm$^{-3}$ for large positive ions. Although ion concentrations in each category fall, time-variations in one category need not exactly follow the variations in other category. Air-earth current density also falls to very low or almost zero value during the period of snowfall.

### 3.4. Effect of blizzard

Simpson [1919] observed that blizzards are intensely electrified and produce high positive potential gradients on the ground. In our observations we observe that whenever high winds are accompanied with some snowfall, i.e., atmospheric temperature are below freezing point, positive ion concentration of all the three categories begin to decrease about 3 – 4 hours before the appearance of snow. For example, on February 15, 2005, winds begin to strengthen at 1800 UT and snowfall started from 2255 UT. Concentrations of all the three ion categories begin to decrease at 1830 UT and decrease by approximately 2 - 4 times by 2200 UT (Figure 10). Although blizzard continued for the next two days, our observations could not be continued beyond midnight on February 15 when wind speed exceeded 30 ms$^{-1}$.

**3.5. Effect of snow-covered ground**

α- and β- rays emitted from the ground are primary sources of ionization close to ground. α- rays are emitted by the radioactive decay of radon. The soil around Maitri is found to have an uranium ($U^{238}$) content varying from 0.036 to 0.364 ppm which is very small in comparison to the range of levels of 1.47 to 4.07 ppm estimated from various types of Indian soil [Ramachandran and Balani, 1995]. In the region free of human activity, such as at Maitri, presence of $U^{238}$ in the soil is mainly attributed to geochemical processes. These emissions can be effectively trapped by a thin layer of ice over ground. During the period of our observations, ample opportunity was provided when the earth's surface at Maitri was completely bare or covered with a layer of snow. The snow that deposited on the ground during the snowfall on January 22 - 23, 2005 completely covered the ground but melted away and left the ground bare by January, 25, 2005. Figure 11 shows the positive ion concentrations and air-earth current density on January 23 and 25, 2005 when the ground was completely covered or bare, respectively. Both days were otherwise marked with fair weather and bright sunshine. Concentrations of all categories of positive ions increase by 100 to 400% on January 25. An important feature of observations is almost uniform concentration of about $3 \times 10^2$ ions cm$^{-3}$ of small positive ions throughout the day on January 23 when the ground is covered with snow.

## 4. Discussion

The small ions produced in the atmosphere soon get attached to aerosol particles and form large ions. In the absence of any other mechanisms directly producing large ions, the variations in small and large ion concentrations are therefore expected to closely follow each other. The close parallelism observed here between the small- and large- positive ion concentrations during the fair-eather periods strongly supports this fact that the ion-aerosol attachment process is the only dominant source for the production of large ions at Maitri. It also demonstrates almost total absence of any effect due to the ions or aerosols produced from anthropogenic sources on land on the Antarctic measurements. Further, almost parallel variations in the air-earth current density and small/large positive ion concentrations confirm the fact that the air-earth current is mainly determined by the small ions. The electric field measurements made in 1997 in the same season at this station show that its diurnal variation exhibits a maximum at 1300 UT and another secondary maximum at 1900 UT with its daily average value as 83 Vm$^{-1}$ [Deshpande and Kamra, 2001]. During the same season, the mean total conductivity value is $2.1 \times 10^{-14}$ Sm$^{-1}$ [Deshpande and Kamra, 2001].

Large ion concentrations in our observations are certainly higher than the expected at a clean site, like Maitri. Some of the likely reasons for this observation are: (i) Being a costal station, some of the marine aerosols being generated over ocean may be transported to the site; this being specially so during the periods when the cyclonic storms revolving around the Antarctic continent, penetrate up to the Maitri station, which happens quite frequently, (ii) Some of the intermediate ions being generated over ice glacier, south of the Maitri grow to the large-ion size during their transportation to Maitri with the prevailing wind from southeast, (iii) Ion counter is placed with its inlet 60 cm above the ground which is bare in this season at Maitri. High winds prevailing at Maitri can strip-off some dust particles from the sandy and loamy sand type of soil. Such dust particles, through expected to be large in size, may carry multiple charges so as to be counted as large ions.

Another factor that may contribute to large ion concentration is the formation of some doubly and multiply charged aerosols during ion-aerosol interactions. These multiply charged aerosol particles even though larger in size as compared to large ions can be collected in large ion condenser due to their higher mobilities and be counted as large ions. The percentage of multiply charged ions in the aerosols plays a significant role only in very pure air having aerosol concentration of < 5000 $cm^{-3}$ [Hogg, 1934 a, b]. Moreover, multiply charged ions gain in importance only when radius of the particles exceeds 10 nm [Israel, 1970; Fuchs, 1964]. The environmental conditions at Antarctica where aerosol concentrations are low and aerosols generally consist of aged particles are suitable for the formation of such multiply charged ions. However, percentage of such multiply charged ions is not likely to exceed 20 % of singly charged ions [Hogg, 1934 a, b; Fuchs, 1964] and is not included in our calculations.

On the contrary, almost independent behavior of the variations of intermediate positive ion concentrations at Maitri suggests that the mechanism responsible for their production is independent of the mechanism responsible for small/large positive ion production. Moreover, the observations that the variations in intermediate positive ion concentration and atmospheric temperature are almost parallel to each other on the fair-weather days and this parallel behavior between the two parameters disappears on cloudy days, strongly suggest that the mechanism responsible for the production of intermediate positive ions is strongly dependent on the solar radiation. Supporting this is the fact that inspite of different surface conditions on the two days in Figure 11 the diurnal variations in the intermediate positive ion concentration and surface temperature on both days are similar because both are fair weather days and have plenty of solar radiation throughout the day. With the abundance of solar radiation, sulphate and low humidity conditions present in the Antarctic environment the production of intermediate ions can occur by nucleation process. Horrak et al. [1998] suggested that cluster ions can grow into intermediate size ions through ion-induced nucleation processes i.e. accumulation of environmental vapours under proper conditions.

The decrease observed in all types of positive ion concentrations and the air-earth current density during snowfall indicates that snow particles effectively scavenge the ions. The fact that the air-earth current density almost reduces to zero value indicates that scavenging of atmospheric ions is almost total at that time. The observations that the decrease in different ion categories is not always parallel to each other are likely to result from the non-uniform rate of scavenging of the ions of different sizes.

The observed decrease in all categories of positive ion concentrations when the ground is covered by a thin layer of snow shows an effective trapping of α and β rays from the ground by a thin layer of snow. Though small but rather distinct difference in small ion concentration over a snow-covered or bare ground supports the observations of Ramachandran and Balani [1995] that radioactive emissions from the rocks at Schirmacher are low and almost comparable to that of sea water. Further, almost constant concentration of small positive ions throughout the whole day when the ground is covered with snow indicates almost uniform production of small ions by cosmic rays and $\gamma$ rays in the lower atmosphere.

**Acknowledgement**

Authors express their gratitude to National Centre for Antarctic and Ocean Research (NCAOR) and Department of Ocean Development (DOD) for participation in the 24$^{th}$ Indian Antarctic Expedition. The meteorological data provided by India Meteorological Department is thankfully acknowledged. (DS) would like to thanks to DST, Govt. of India, under the BOYSCAST programme (with reference SR/BY/A-19/05).

**References**

Bering, E. A., A. A. Few, and J. R. Benbrook (1998), The global electric circuit, Phys. Today (Oct. issue), 24-30.

Byrne G. J., J. R. Benbrook, E. A. Bering, A. A. Few, G. A. Morris, W. J. Trabucco, and E. W. Paschal (1993), Ground-based instrumentation for measurements of atmospheric conduction current and electric-field at the south-pole, J.Geophys. Res., 98, D2, 2611-2618.

Burns, G. B., M. H. Hesse, S. K. Parcell, S. Malachowski and K. D. Cole (1995), The geoelectric field at Devis station, Antarctica, J. Atmos. Terr. Phys., 57, 1783-1797.

Carslaw, K. S., R. G. Harrison, and J. Kirkby (2002), Cosmic rays, clouds, and climate, Science, 298, 1732-1737.

Cobb, W. E., and H. J. Wells (1970), The electrical conductivity of oceanic air and its correlation to global atmospheric pollution, J. Atmos. Sci. 27, 814–819.

Deshpande, C. G., and A. K. Kamra (2001), Diurnal variations of atmospheric electric field and conductivity at Maitri, Antarctica, J. Geophys. Res., 106, 14207-14218.

Deshpande, C. G., and A. K. Kamra (2002), Atmospheric electric conductivity measurements over the Indian Ocean during the Antarctic Expedition in 1996-1997, J. Geophys. Res., 107, doi: 10.1029/2002JD002118.

Dhanorkar, S., and A. K. Kamra (1991), Measurement of mobility spectrum and concentration of all atmospheric ions with a single apparatus, J. Geophys. Res., 96, 18671-18678.

Dhanorkar, S., and A. K. Kamra (1992), Relation between electrical conductivity and small ions in the presence of intermediate and large ions in the lower atmosphere, J. Geophys. Res., 97, 20345-20360.

Dhanorkar, S., and A. K. Kamra (1993), Diurnal variations of the mobility spectrum of ions and size distribution of fine aerosols in the atmosphere, J. Geophys. Res., 98, 2639-2650.

Fuchs, N. A. (1964), The mechanics of aerosols, Pergamon Press Inc., New York, pp. 408.

Harrison, R. G., and K. S. Carslaw (2003), Ion-aerosol-cloud processes in the lower atmosphere, Rev. Geophys., 41, doi: 10.1029/2002RG000114.

Hirsikko, A., Laakso, L., Urmas, H., Aalto, P. P., Kerminen, V. M., and Kulmala, M. (2005), Annual and size dependent variation of growthrates and ion concentrations in boreal forest. Boreal Env. Res. 10, 357-369.

Hogg, A. R. (1934 a), Atmospheric Electric Observations, Gerl. Beitr. Geophys., 41, 1-31.

Hogg, A. R. (1934 b), Some observations of the average life of small ions and atmospheric ionization equilibria, Gerl. Beitr. Geophys., 41, 32-54.

Horrak, U., J. Salm, and H. Tammet (1998), Bursts of intermediate ions in atmospheric air, J. Geophys. Res., 103, 13909-13915.

Horrak, U., J. Salm, and H. Tammet (2000), Statistical characterisation of air ion mobility spectra at Tahkuse observatory: Classification of air ions, J. Geophys. Res., 105, 9291–9302.

Horrak, U., J. Salm, and H. Tammet (2003), Diurnal variation in the concentration of air ions of different mobility classes in a rural area, J. Geophys. Res., 108, doi: 10.1029/2002JD003240.

Israel, H. (1970), Atmospheric Electricity, Vol. I, Israel Program for Scientific Translations Ltd., The National Science Foundation, Washington, D. C., pp. 317.

Jonassen, J., and M. H. Wilkening (1965), Conductivity and concentration of small ions in the lower atmosphere, J. Geophys. Res., 70, 779-784.

Kamra, A. K., and C. G. Deshpande (1995). Possible secular change and land–to-ocean extension of air pollution from measurements of atmospheric electrical conductivity over the Bay of Bangal, J. Geophys. Res., 100, 7105–7110.

Kasemir, H. W. (1972), Atmospheric electric measurements in the Arctic and Antarctic, Pure Appl. Geophys, 100, 70–80.

Markson, R. (1978), Solar modulation of atmospheric electrification and possible implication for the Sun-Weather relationship, Nature, 273, 103–105

Markson, R. and M. Muir (1980), Solar wind control of the Earth's electric field, Science, 208, 979–990.

Misaki, M. (1961), Studies on the atmospheric ion spectrum, II, Relation between the ion spectrum and the electrical conductivity, Pap. Meteorol. Geophys. Tokyo, 12, 261-276.

Misaki, M., M. Ohtagaki, and I. Kanazawa (1972), Mobility spectrometry of the atmospheric ions in relation to atmospheric pollution, Pure Appl. Geophys. 100, 133-145.

Nadykto, A., and Yu, F. (2003), Uptake of neutral polar vapour molecules by charged particles: Enhancement due to dipole-charge interaction, J. Geophy. Res., 108(D23), 10.1029/2003JD003664.

Norinder, H., and R. Siksna (1953), Variations of the concentration of ions at different heights near the ground during quiet summer nights at Uppsala, Ark. Geophys, 1, 519-541.

Ramachandran, T. V., and M. C. Balani (1995), Report on the participation by the Bhabha Atmomic Research Centre in the Tenth Indian Expedition to Antarctica, Scientific Report on Tenth Indian Expedition to Antarctica, Tech. Publ. No. 8, Dept. of Ocean Development, Govt. of India, 159-180.

Roble, R. G. and I. Tzur, (1986), The global atmosphere electrical circuit, Study in Geophysics -The Earth electrical Environment, National Academy Press, Washington, D.C., 206–231.

Rycroft, M. J., S. Israelsson, and C. Price (2000), The global atmospheric electric circuit, Solar Activity and Climate Change, J. Atmos. Solar Terr. Phys., 62, 1563–1576.

Siingh, Devendraa, V. Gopalakrishnan, R. P. Singh, A. K. Kamra, Shubha Singh, Vimlesh Pant, R. Singh, and A. K. Singh (2007), The atmospheric global electric circuit: An overview, Atmos. Res., 84, 91-110.

Simpson G. C., (1919), British Antarctic Expedition 1910–1913. Thacker, Spink and Co., 326 pp.

Singh, D. K., R. P. Singh, and A. K. Kamra, (2004), The electrical environment of the Earth's atmosphere: a review, Space Sci. Rev., 113, 375-408.

Tinsley, B. A. (2000), Influence of solar wind on the global electric circuit, and inferred effects on the cloud microphysics, temperature, and dynamics in troposphere, Space Sci. Rev., 94, 231–258.

Tinsley, B. A. and R. A. Heelis (1993), Correlations of atmospheric dynamics with solar activity. Evidence for a connection via the solar wind, atmospheric electricity and cloud microphysics, J. Geophys. Res. 98, 10375–10384.

Virkkula, A., Vana, M., Hirsikko, A., Aalto, P. A., Kulmala, M., and Hillamo, R. (2005), Air ion mobility and aerosol particle size distributions at the Finnish Antarctic research station Aboa, Proceedings of European Aerosol Conference 2005, Ghent, Belgium.

Wait, G. R., and D. W. Torreson (1934), The large-ion and small-ion content of atmosphere at Washington, D. C., J. Geophys. Res., 39, 111-119.

**Legends**

Figure 1. Ion counter installed at Maitri

Figure 2. Air-earth current plate antenna installed at Maitri.

Figure 3. The map of Antarctica, showing the location of Maitri station. Location of instruments (Kamet observatory) at Maitri and the wind rose showing the magnitudes and directions of wind speed during the January to February 2005 period.

Figure 4. Daily average values of the small-, intermediate-, and large- positive ion concentrations and air-earth current density for Julian days in 2005. Vertical bars show the standard deviations.

Figure 5. Daily average values of meteorological parameters.

Figure 6. Variations in small-. intermediate-, and large- positive ion concentrations and air-earth current density on January 12 (panel a) and January 3 (panel b), 2005.

Figure 7. Diurnal variations of small-, intermediate- and large- positive ion concentrations and air-earth current density on a fair-weather day, January 10, 2005.

Figure 8. Diurnal variations of small-, intermediate and large- positive ion concentrations and air-earth current density on a cloudy day, January 18, 2005.

Figure 9. Change in concentrations of positive ions and air-earth current density due to snowfall on January 22, 2005.

Figure 10. Change in concentrations of positive ions and air-earth current density during a blizzard on February 15, 2005.

Figure 11. Diurnal variations of the small-, intermediate- and large- positive ion concentrations and air-earth current density on two fair-weather days when the ground was covered with snow on January 23, 2005 (a) or was bare on January 25, 2005 (b).

**Table 1:** Dimensions and other parameters of three condensers of the ion-counter.

| Dimensions/constants | Small-ion Condenser | Intermediate-ion Condenser | Large-ion Condenser |
|---|---|---|---|
| Length of the outer electrode (m) | 0.4 | 0.8 | 1.2 |
| Length of the inner electrode (m) | 0.2 | 0.5 | 1.0 |
| Diameter of the outer electrode (m) | 0.098 | 0.06 | 0.038 |
| Diameter of the inner electrode (m) | 0.076 | 0.037 | .022 |
| Potential applied (V) | 15 | 100 | 600 |
| Critical mobility ($m^2 V^{-1} s^{-1}$) | $0.766 \times 10^{-4}$ | $1.2 \times 10^{-6}$ | $0.97 \times 10^{-8}$ |
| Flow rate ($l\ s^{-1}$) | 8.6 | 1.8 | 0.29 |

**Table 2:** The mobility and size ranges of ions

| Category | Small ions | Intermediate ions | Large Ions |
|---|---|---|---|
| Mobility range ($m^2\ V^{-1}\ s^{-1}$) | $> 0.77 \times 10^{-4}$ | $1.21 \times 10^{-6} - 0.77 \times 10^{-4}$ | $0.97 \times 10^{-8} - 1.21 \times 10^{-6}$ |
| Diameter range (nm) | < 1.45 | 1.45 – 12.68 | 12.68 - ~130 |

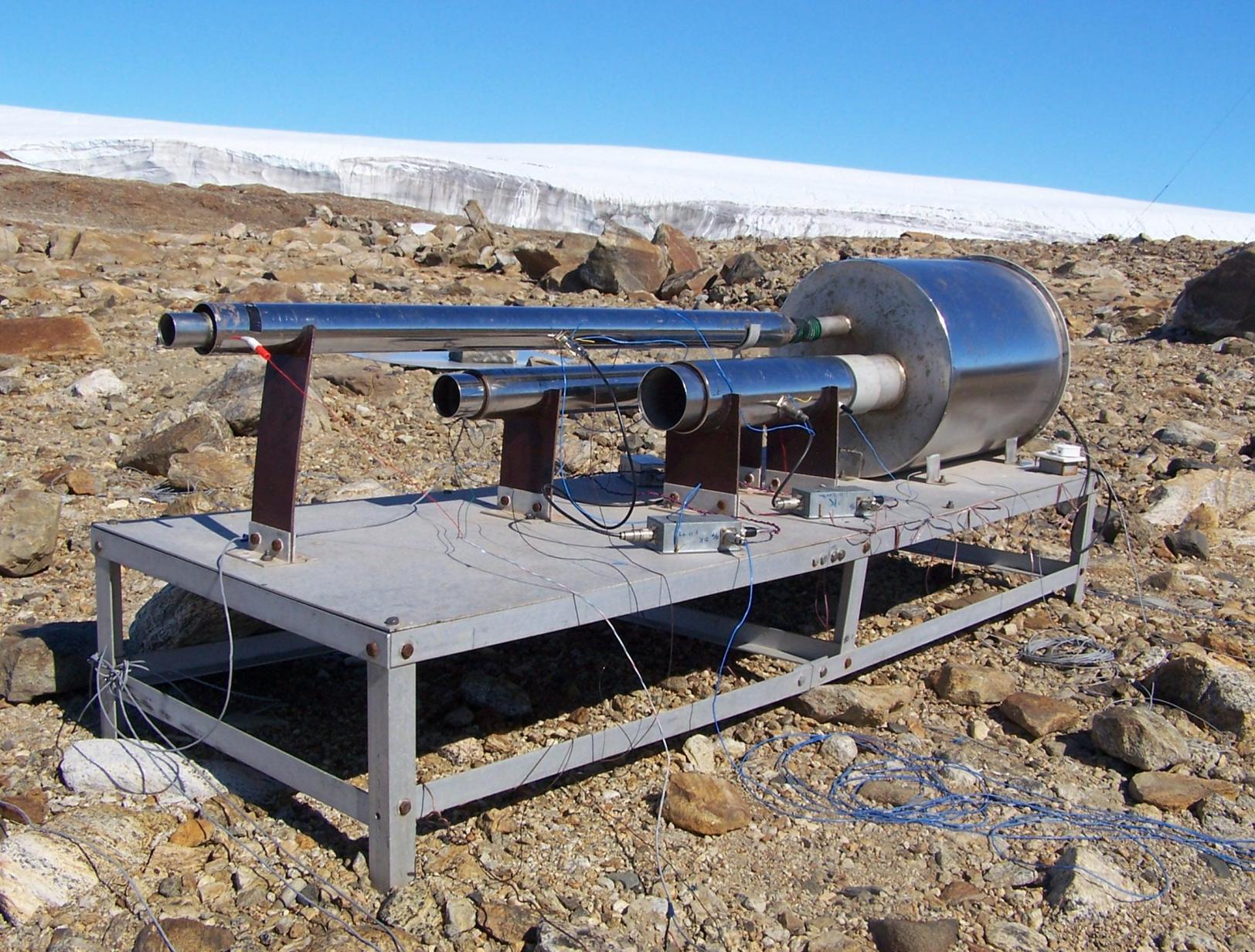

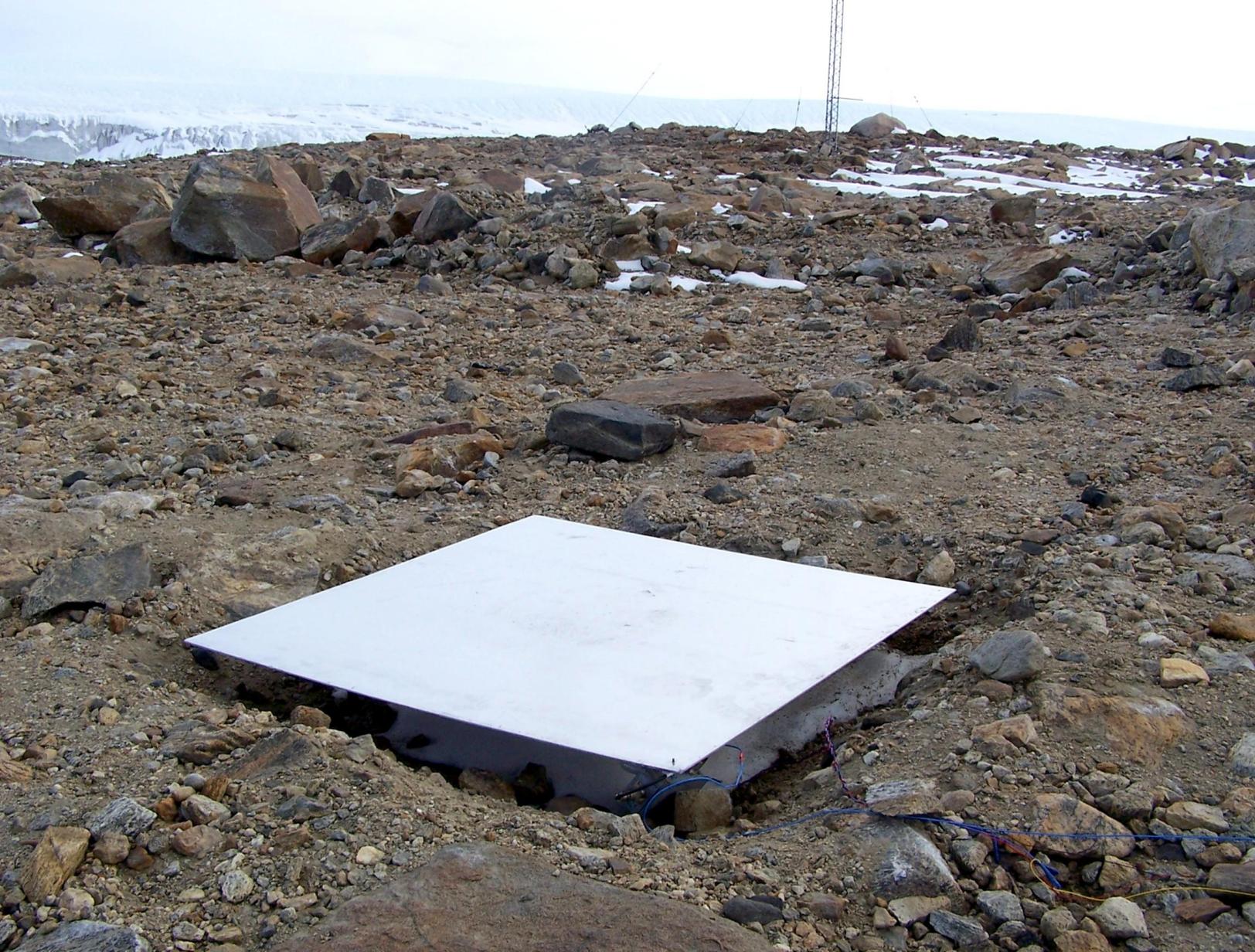

Figure 3

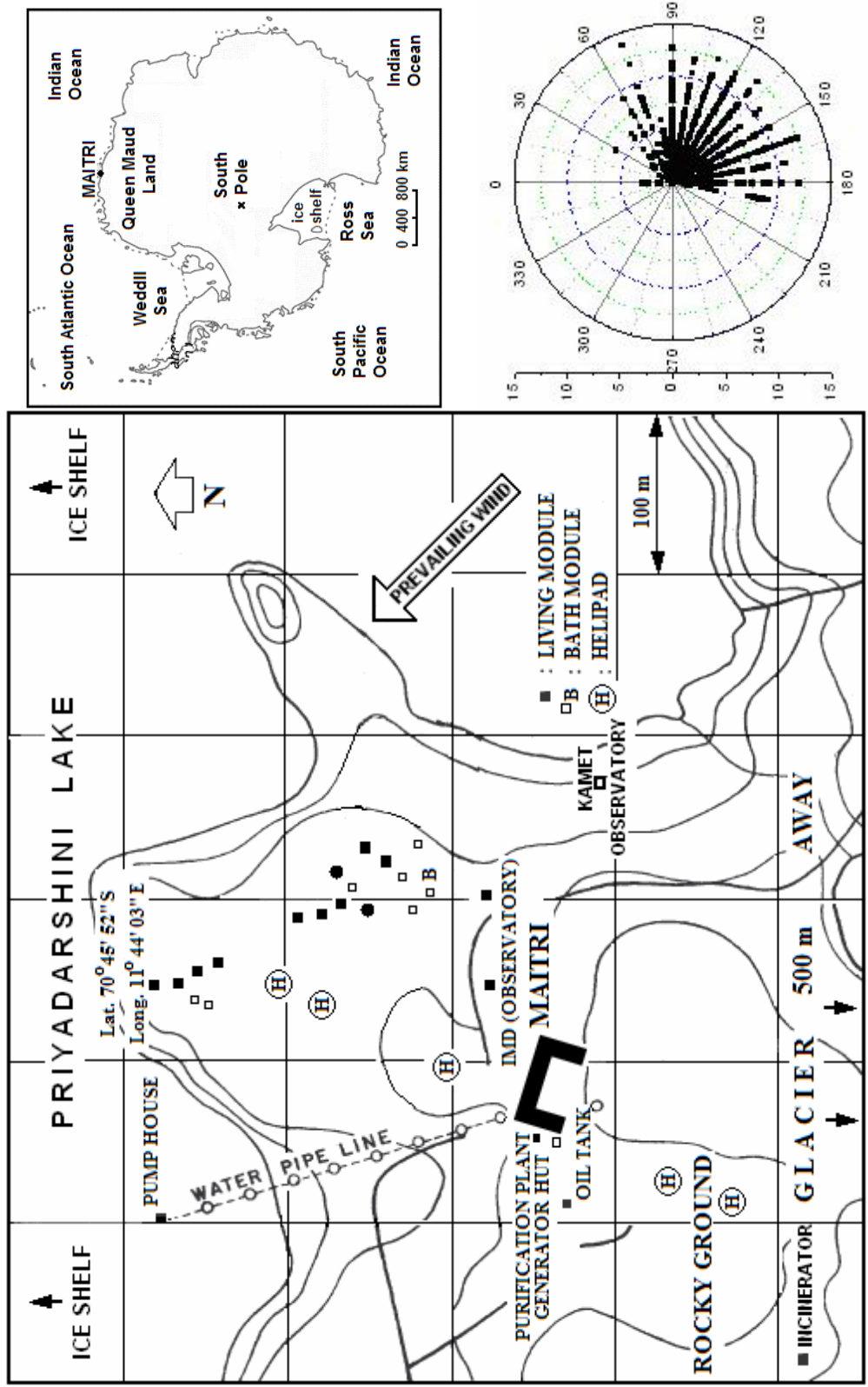

Figure 4

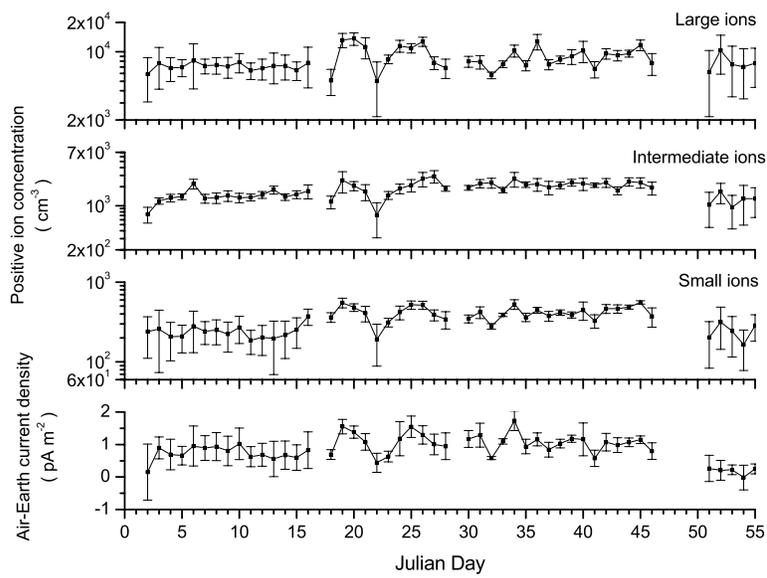

Figure 5

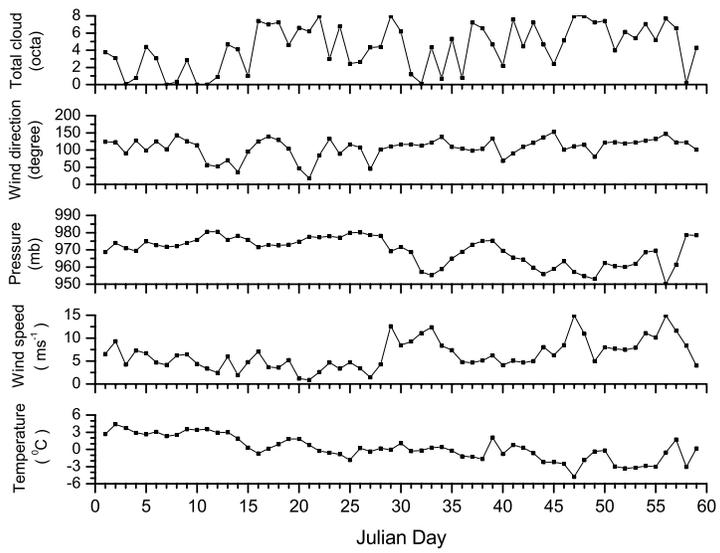

Figure 6

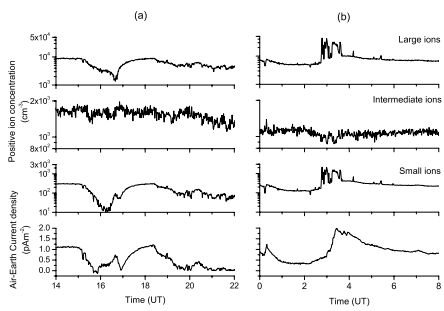

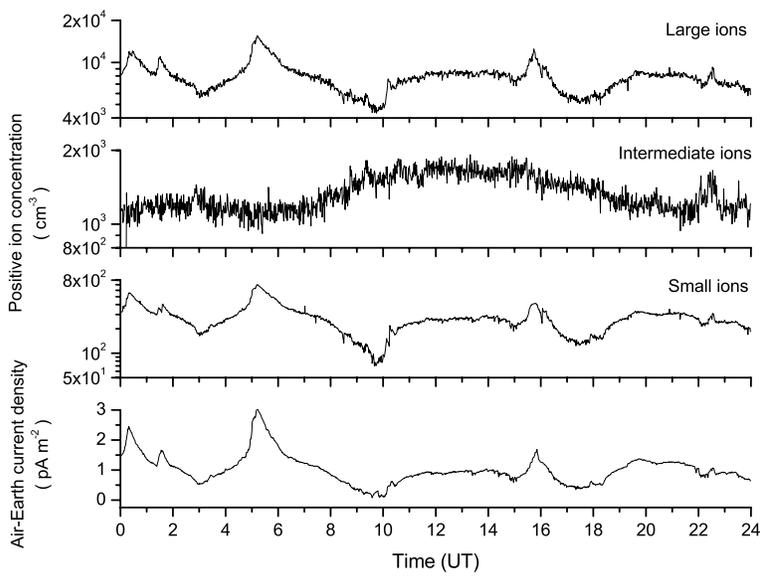

Figure 7

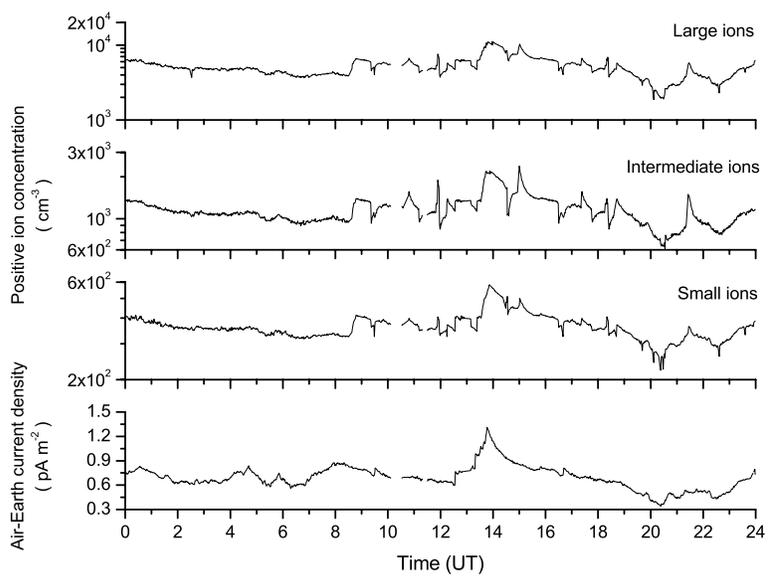

Figure 8

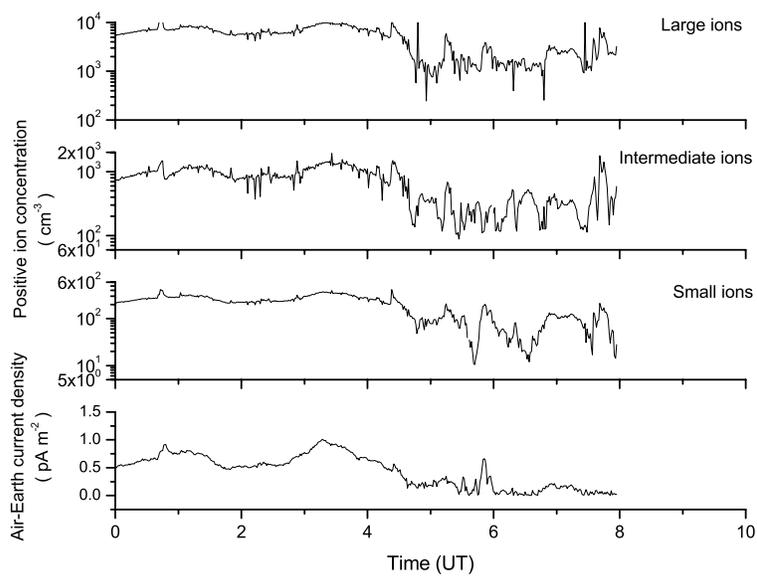

Figure 9

Figure 10

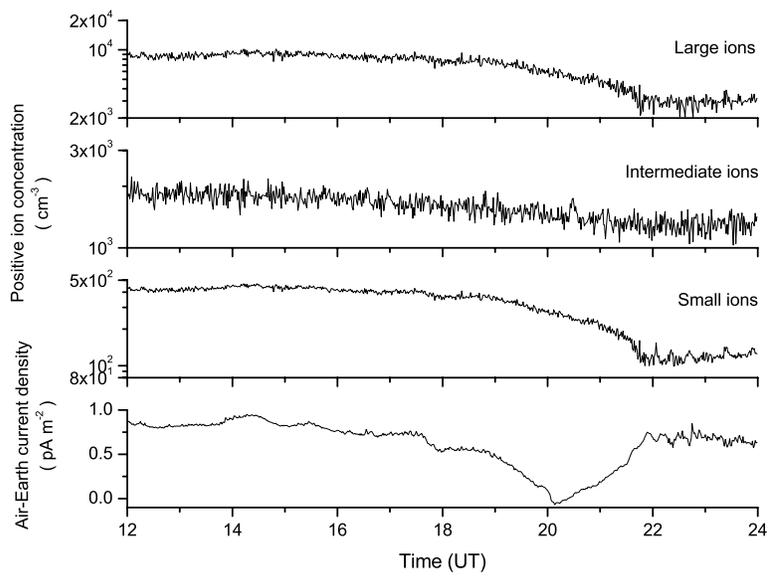

Figure 11

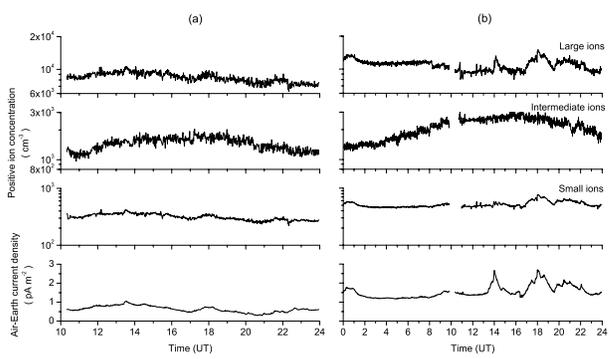